\documentclass{article}
\usepackage{amsmath}
\usepackage{graphicx}
\usepackage{amssymb}
\usepackage{geometry}

\def\proba{\text{Proba}}
\def\diffd{\text{d}}

\title{How genealogies are affected by the speed of evolution}

\date{\today}

\author{\'Eric Brunet\footnote{Email: \texttt{Eric.Brunet@lps.ens.fr}}\ and
Bernard Derrida\footnote{Email: \texttt{Bernard.Derrida@lps.ens.fr}}\\[6pt]
Laboratoire de Physique Statistique, \'Ecole Normale Sup\'erieure, \\
UPMC, Universit\'e Paris Diderot, CNRS,\\
 24 rue Lhomond, 75231 Paris cedex 05, France
}

\begin{document}
\maketitle
 
\begin{abstract}
In a series of recent works it has been shown that a class of simple
models of evolving populations under selection leads to genealogical
trees whose statistics are given by the Bolthausen-Sznitman coalescent
rather than by the well known Kingman coalescent in the case of neutral
evolution. Here we show that when conditioning the genealogies on the
speed of evolution, one finds a one parameter family of tree statistics
which interpolates between the Bolthausen-Sznitman and Kingman's
coalescents. This interpolation can be calculated explicitly for one
specific version of the model, the exponential model. Numerical
simulations of another version of the model and a phenomenological theory
indicate that this
one-parameter family of tree statistics could be universal. We compare
this tree structure with those appearing in other contexts, in particular
in the mean field theory of spin glasses.

\end{abstract}

\section{Introduction}

An important question in the study of evolving populations is to
understand the effect of
selection on the ancestry and on the genealogies \cite{HudsonKaplan.88,
Hudson.91, NeuhauserKrone.97, Nordborg.01, DegnanSalter.05}. In absence
of selection, for a well mixed population such as in the Wright-Fisher
model, the statistical properties of the genealogy of a large population
of constant size is described by Kingman's coalescent \cite{Kingman.82,
Kingman2.82, DonnellyTavare.95, DerridaPeliti.91}. Recent attempts to
modify the Wright-Fisher model in order to introduce selection lead to
a change of the statistical properties of genealogies: in \cite{BDMM2.06,
BDMM.07}, the
study of a whole class of models indicates that the genealogies of
populations evolving under selection are given by Bolthausen-Sznitman's
coalescent \cite{BolthausenSznitman.98} rather than by Kingman's (this has been shown
analytically only for one specific version of the model, the exponential model (see
below), but it has also been checked in numerical simulations and proved
for a modified version of the model where the effect of selection is
represented by a moving absorbing wall along the fitness axis
\cite{BerestyckiBerestyckiSchweinsberg.10}).
In the present paper, we further study these simple models of evolution
with selection and we calculate how the statistical properties of the
genealogies are correlated to the speed of evolution.

\begin{figure}[ht]
\centering
\includegraphics[width=.5\textwidth]{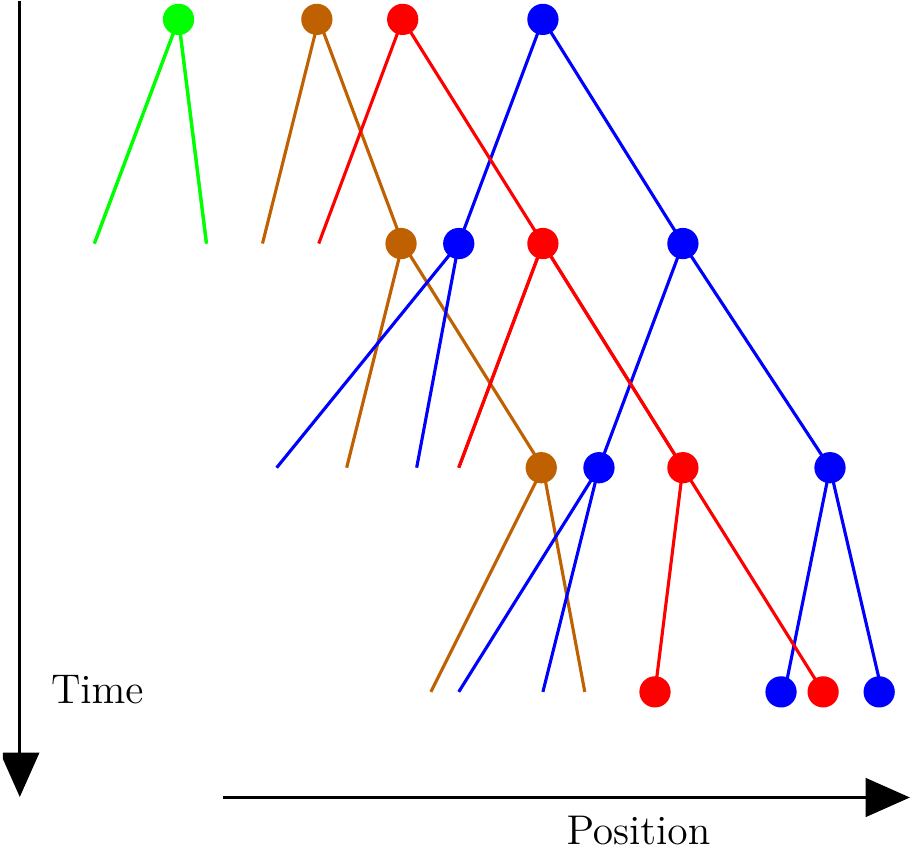}
\caption{Three time steps for a population of size $N=4$ in a model where each individual has
two offspring.}
\label{fig:ex}
\end{figure}

The models of evolution with selection we consider here have been
introduced in \cite{BDMM2.06,BDMM.07} (see also \cite{Snyder.03, DurrettRemenik.09}). They can be
defined as follows: at each generation $t$ the population consists of
a fixed number $N$ of individuals and each individual $i$ is characterized
by a single number $x_i(t)$ representing its adaptation in the environment.
So $x_i(t)$ is the position of individual $i$ on a fitness or adaptation
axis (very much like in the Bak-Sneppen model \cite{BakSneppen.93}). This individual
has several offspring at positions $x_i(t)+\epsilon_{i,1}(t)$,
$x_i(t)+\epsilon_{i,2}(t)$, $x_i(t)+\epsilon_{i,3}(t)$, etc., where the
$\epsilon_{i,j}(t)$ are random numbers representing the change of
adaptation due to mutations between
parent $i$ and child $j$. The total number of offspring produced this way
by all individuals at a given generation $t$ exceeds $N$; the population at
generation $t+1$ is then obtained by keeping the $N$ most adapted
children (i.e. the $N$ rightmost points along the axis) among all these
offspring, see figure~\ref{fig:ex}. 
The model is fully specified when the distribution of the number offspring
of each individual and the distribution of the random
shifts $\epsilon$ are given.

After letting such a model evolve for a large number $t$ of generations, the
positions $x_i(t)$ of the individuals on the adaptation axis from a cloud of points 
grouped around a position $X_t$ which grows linearly with time
with some velocity $v_N$. There is some arbitrariness in the way this
position $ X_t$ can be defined (one could choose for example $X_t$ to be
the position of the rightmost individual, or of the leftmost individual
or the center of mass of the population) but, as the positions of all the
individuals remain grouped, a change of definition modifies the value of
$X_t$ by an amount which does not grow with time and therefore does not
affect the velocity $v_N$. In addition to the velocity, the position $X_t$
has fluctuations: in particular it diffuses with a variance which grows
linearly in time \cite{BDMM.06}.

\begin{figure}[ht]
\centering
\includegraphics[width=.7\textwidth]{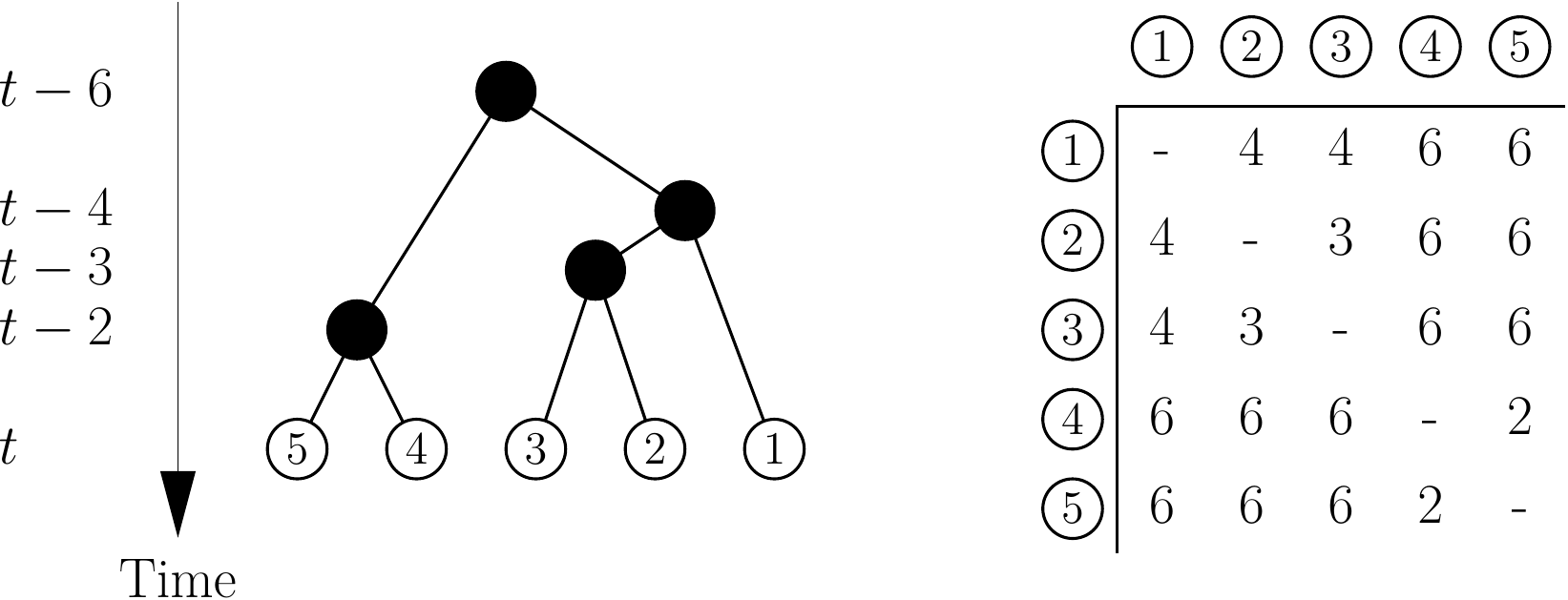}
\caption{\emph{Left:} an exemple of the genealogical tree of $N=5$
individuals.\quad\emph{Right:} the coalescence times $\tau_{i,j}(t)$
corresponding to the tree.}
\label{fig:coltimes}
\end{figure}

At each generation, one can study the genealogy of the population of this
model by considering the matrix $\tau_{i,j}(t)$ of the ages of the most recent common
ancestors, or \emph{coalescence times}, of all pairs of individuals $i$ and $j$ living at
generation $t$; see figure~\ref{fig:coltimes}. As the genealogy is a
tree, the whole ancestry of the population can be deduced from the
knowledge of this matrix. In particular the age of the most recent common
ancestor of any subset of the population can be expressed in terms of
this matrix: for example the age $\tau_{i,j,k}(t)$ of
the most recent common ancestor of three individuals $i$, $j$ and $k$ at
generation $t$ is simply
\begin{equation}
\tau_{i,j,k}(t)=\max\big(\tau_{i,j}(t),\tau_{i,k}(t),\tau_{j,k}(t)\big).
\end{equation}

One property common to the models of evolution under selection studied here
and to the neutral models of evolution described by Kingman's coalescent is
that the heights and the shapes of the genealogical trees
fluctuate with $t$ even when
the size $N$ of the population becomes large. The statistical properties of these trees 
and their time scales are however different: for
instance, the
typical age of the most recent common ancestors of $k$ individuals grows
logarithmically with the population size~$N$ in presence of selection (as in
the models studied here) while it grows linearly with $N$ in the neutral
case.

There are several ways of describing the statistical properties of these
trees (see section~\ref{sec:poissondirichlet}). In \cite{BDMM2.06,BDMM.07}
we chose to characterize them by the average coalescence times $\langle T_k
\rangle $ of $k$ individuals chosen at random in the population:
\begin{equation}
\langle T_k \rangle = \big\langle \tau_{i_1,\ldots,i_k}(t)\big\rangle.
\end{equation}
(here $\langle \cdot \rangle$ denotes an average
over the individuals $i_1,\ldots,i_k$ and over the generation $t$).
The $N$ dependence of $\langle T_2\rangle$ gives the time scale over which coalescence events 
occur, while the ratios $\langle T_k\rangle/\langle T_2\rangle$ are a signature of the statistical properties of 
the shape of the trees. We found in \cite{BDMM2.06,BDMM.07} that these ratios in
presence of selection converge when $N\to\infty$ to those of
a Bolthausen-Snitzmann coalescent:
\begin{align}
&\frac{\langle T_3 \rangle }{ \langle
T_2 \rangle} = \frac{5 }{ 4},\qquad
\frac{\langle T_4 \rangle}{ \langle T_2 \rangle}
 = \frac{25 }{ 18},&&\text{(Bolthausen-Snitzmann)}\label{selection}\\
\intertext{in contrast to the neutral case where they converge to those of Kingman's
coalescent:}
&\frac{\langle T_3 \rangle }{ \langle T_2 \rangle} =
\frac{4 }{ 3} , \qquad \frac{\langle T_4 \rangle }{ \langle T_2 \rangle} =
\frac{3 }{ 2}.&&\text{(Kingman)}
\label{neutral}
\end{align}

Our goal here is to calculate how these ratios are correlated to the
speed of evolution, by weighting all the events during a long time
interval $\tau$ by a factor $e^{-\beta X_\tau}$. ($\beta<0$ favors events with
a speed of evolution faster than average, while $\beta>0$ correponds to
events with a slower speed of evolution.) Our main result, derived below
for the exponential model, is that the above ratios become for large~$N$
\begin{equation}
\frac{\langle T_3 \rangle }{ \langle T_2 \rangle} 
	= \frac{5+4 \beta }{ 4+3 \beta} , \qquad 
\frac{\langle T_4 \rangle}{ \langle T_2 \rangle} 
	= \frac{100+204 \beta+ 133 \beta^2+ 27 \beta^3}
		{72 + 142 \beta+ 90 \beta^2 + 18 \beta^3}
\label{beta}
\end{equation}
It is remarkable that these expressions interpolate between the neutral
case (Kingman) for $\beta\to +\infty$ (low speed limit)
and the selection case (Bolthausen-Snitzmann) for $\beta=0$. When
$\beta\to-1$ (high speed
limit), all the ratios become $1$ indicating a ``star-shaped'' coalescent.

This paper is organized as follows: in section~\ref{sec:weight}, we
explain how the weighting by the factor $e^{-\beta X_\tau}$ is done. In
section~\ref{sec:exp}, we show that in presence of the bias, one version
of the model (the exponential model) can be solved exactly by analyzing a
coalescent model, the rates of which depend on $\beta$. This leads to
(\ref{beta}). In section~\ref{sec:phenom} we argue using the
phenomenological theory developed in \cite{BDMM.07} that \eqref{beta}
should remain valid for other versions of the model up to a change of
scale of $\beta$. Lastly in section~\ref{sec:poissondirichlet} we compare
the $\beta$-random tree structure which leads to \eqref{beta}
to the statistics of the partitions in
mean field spin glasses and in the Poisson-Dirichlet distribution.

\section{How to condition on the velocity}\label{sec:weight}

If one performs a simulation of the model described in the introduction,
one can measure at each generation $t$ the position $X_t$ of the
population (defined in any reasonable way: as explained in the
introduction, the precise definition does not matter) and the ages $T_2(t),
\ldots, T_k(t)$ of the most recent common ancestor of $2,\ldots,k$ individuals
chosen at random in the population at time~$t$ (for more efficiency one
can average these times $T_k(t)$ over all the choices of the $k$
individuals in the population at time~$t$). Then we choose a long time
interval $\tau$ and we want to determine 
\begin{equation}
\langle T_k \rangle_\beta = \lim_{\tau \to \infty} \frac1\tau
\sum_{t=1}^\tau
\frac{\big\langle T_k(t) e^{-\beta X_\tau} \big\rangle} {\big\langle e^{-\beta
X_\tau} \big\rangle }.
\label{Tbeta}
\end{equation}

\subsection{Theoretical considerations}

As the model is a Markov process and correlations decay fast enough in
time, we expect that, for large~$t$,
\begin{equation}
\left\langle e^{-\beta (X_t-X_0)} \right\rangle \sim e^{ t G(\beta)}.
\label{LD}
\end{equation}
The knowledge of $G(\beta)$ determines all the cumulants of the position
$X_t$
\begin{equation}
 \lim_{t \to \infty}\frac{\langle X_t^n \rangle_c} t 
 = (-)^n \left.\frac{\diffd^n G(\beta)} { \diffd \beta^n}
 \right|_{\beta=0}.
\end{equation}
It is also related to the large deviation function $F(v)$ of the velocity
defined by
\begin{equation}
\proba(X_t = v t) \sim e^{t F(v)}
\end{equation}
through a Legendre transform
\begin{equation}
G(\beta) = \max_v [-\beta v + F(v)].
\end{equation}
For large $t$, the value of $v$ which dominates the weighted averages
in (\ref{Tbeta}) and (\ref{LD}) is given by
\begin{equation}
v= -\frac{\diffd G(\beta)}{\diffd \beta},
\label{vbeta} 
\end{equation}
with fluctuations of order $t^{-1/2}$.
Therefore the weighted averages (\ref{Tbeta}) become equivalent in the
$t \to \infty$ limit to conditioning on the velocity $v$ given by
(\ref{vbeta}). This is, in the present context, the analog of the
well known equivalence of ensembles in statistical physics.

\subsection{In numerical simulations}

Numerically it is difficult to perform averages such as \eqref{Tbeta}
because the events which dominate both the numerator and the denominator
of (\ref{Tbeta}) are rare events. In order to overcome this difficulty,
we use an importance sampling method. We consider a sample periodic in
time of period $\tau$ where $\tau$ is chosen large enough. This means
that the random
shifts $\epsilon_{i,j}(t)$ are periodic in time
($\epsilon_{i,j}(t+\tau)=\epsilon_{i,j}(t)$ for all $i$, $j$ and $t$). With
these periodic $\epsilon_{i,j}(t)$,
the evolution of the system becomes also periodic in time:
the shift $X_\tau$ of the position of the population after one
period~$\tau$ can therefore be unambigously defined and depends on
all the $\epsilon_{i,j}(t)$. Then, we perform a standard Monte-Carlo
simulation: at each step we try a new sample by changing some of
the $\epsilon_{i,j}(t)$ and we let the system evolve till it becomes
periodic (here we change all the $\epsilon_{i,j}(t)$
at a random time $t$ uniformly distributed between $1$ and $\tau $). The
outcome of this change is to modify the $X_\tau$ to a new
value $X_\tau^\text{new}$. Then, as always with a Metropolis algorithm,
we accept the change with a probability $\max \big[1,
\exp[-\beta(X_\tau^\text{new}- X_\tau)] \big]$. With this procedure
samples are produced with a weight $\exp[{-\beta X_\tau}]$, so that by
averaging quantities such as the $T_k$ over many samples one gets an
estimate of \eqref{Tbeta}.

We have simulated the exponential model (see section~\ref{sec:exp} where
we give the precise definition of the exponential model and its analytic
solution in the $N\to\infty$ limit) for $N=100$ and a value of
$\tau\approx 30\ln N$ which is much larger than $\langle T_2\rangle $
\cite{BDMM.07}. For each value of $\beta$, we measured $\langle
T_2\rangle$, $\langle T_3\rangle$, $\langle T_4\rangle$ averaged on
$10^6$ Monte-Carlo steps. We have also simulated a more generic model
where each individual has exactly two offspring  with independent random shifts $\epsilon_{i,j}(t)$ uniformly
distributed between 0 and~1. This model cannot be solved exactly, but a
phenomenological theory (see \cite{BDMM.07} and section~\ref{sec:phenom})
predicts when $N\to\infty$ the same statistics of the genealogical trees \eqref{beta} as in
the exponential model with $\beta$ replaced by $\beta/\gamma$, where
$\gamma\simeq5.262$ is the value which minimizes the function
$\ln[2(e^\gamma-1)/\gamma]/\gamma$, see section~\ref{sec:phenom}. We
simulated the sizes $N=30$, $N=100$ and $N=300$ with values of
$\tau\approx 8\ln^3 N$ which is much larger than $\langle T_2\rangle$,
and again averaged over $10^6$ Monte-Carlo steps. We also checked for
several values of $\beta$ that our results remain unchanged by choosing a
value of the period $\tau$ twice as big (results not shown), indicating
that our Monte-Carlo results would be the same for an infinite
time-window.

\begin{figure}[ht]
\centering
\includegraphics[width=.69\textwidth]{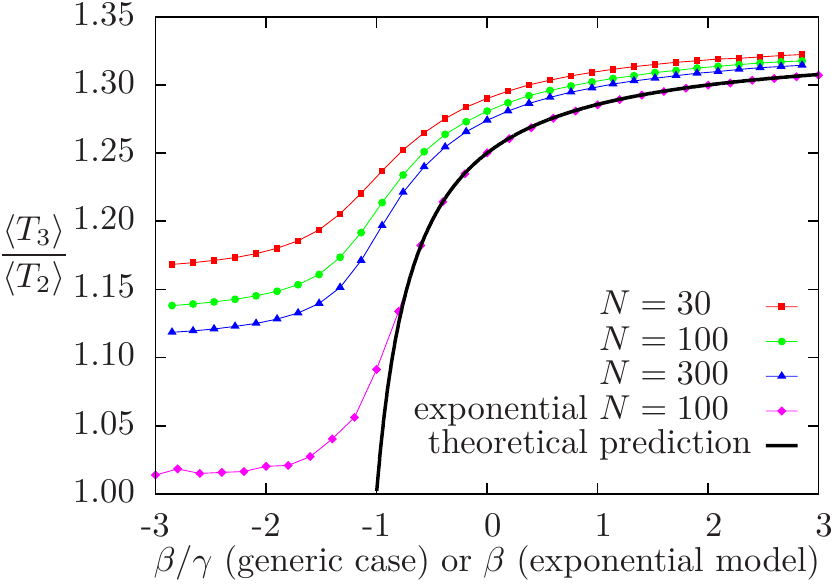}
\caption{$\langle T_3\rangle/\langle T_2\rangle$ as a function of
$\beta$ for the exponential model and the prediction \eqref{beta}, and as
a function of $\beta/\gamma$ for the generic model described in the
text.}
\label{fig:T3}
\end{figure}

\begin{figure}[ht]
\centering
\includegraphics[width=.69\textwidth]{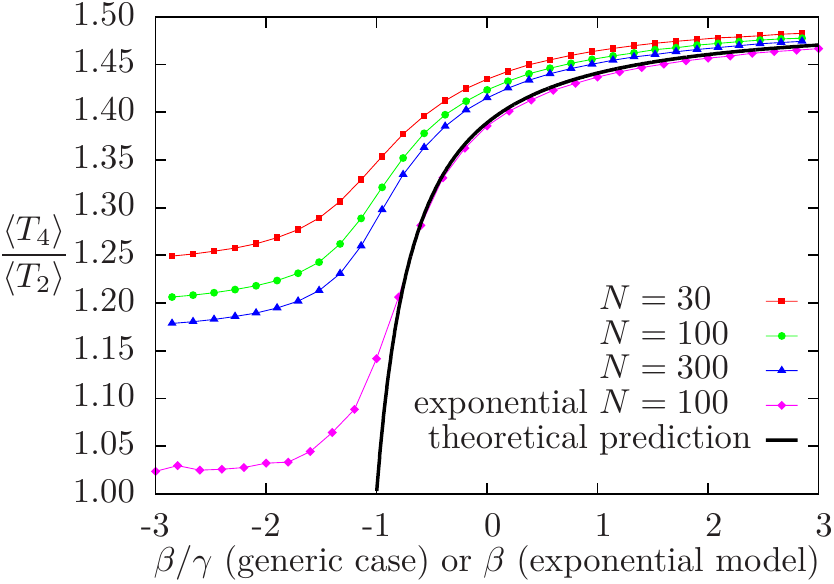}
\caption{$\langle T_4\rangle/\langle T_2\rangle$ as a function of
$\beta$ for the exponential model and the prediction \eqref{beta}, and as
a function of $\beta/\gamma$ for  the generic model described in the
text.}
\label{fig:T4}
\end{figure}

The results for $\langle T_3\rangle/\langle T_2\rangle$ and $\langle
T_4\rangle/\langle T_2\rangle$ are presented in figures~\ref{fig:T3}
and~\ref{fig:T4}. As in \cite{BDMM.07}, we observe that for the
exponential model, the results are already very close  to their
asymptotic limits even for $N=100$; only for $\beta\lesssim-1$ there is a
discrepancy with the theoretical prediction~\eqref{beta}. As in
\cite{BDMM.07}, the convergence is however slower in the generic case but
the curves seem in both cases to converge to the prediction.

\section{Exponential model}\label{sec:exp}

In this section we consider a version of the model, the exponential
model, which can be solved exactly \cite{BDMM.07}. In the exponential
model, the shifts of the offspring of each individual 
are generated by a Poisson process of density $\rho(\epsilon)
= e^{-\epsilon}$. This means that an
individual at position $x_i(t)$ has a probability $e^{-\epsilon} \diffd
\epsilon$ of having an offspring in the infinitesimal interval $(x_i(t)+
\epsilon, x_i(t) + \epsilon + \diffd \epsilon)$. Then the population at
the next generation is obtained by selecting the $N$ rightmost points among
all the offspring produced by generation $t$. (Note that in the exponential
model the number of offspring produced by each generation is infinite
but their number at the right of any position $y$ is finite. There is
therefore no problem to select the $N$ survivors at generation $t+1$).

In the exponential model, there is a convenient way of defining the
position $X_t$ of the population
\begin{equation}
X_t = \ln\bigg[\sum_{i=1}^N e^{x_i(t)}\bigg].
\label{defXt}
\end{equation}
The simplicity of the exponential model comes from the fact that,
with this definition of $X_t$, one has
\begin{equation}
\sum_i e^{-[x-x_i(t)]} = e^{-(x-X_t)},
\end{equation}
which means that one can generate the offspring of the whole population
at time $t$ by replacing the $N$ Poisson processes centered at the
positions $x_i(t)$ by a single Poisson process centered at position
$X_t$. Therefore, with definition (\ref{defXt}) of $X_t$, the $N$ points
$x_i(t+1)$ at generation $t+1$ are the $N$ rightmost points of a Poisson
point process with a density $\exp[{-(x-X_t)}]$.

As explained in \cite{BDMM.07} a way of drawing these $N$ points is to
choose a number $z$ with a density of probability
$\proba(z)=\exp[-(N+1)z-e^{-z}]/N!$ and, independently, $N$ numbers $y_i$
with an exponential density $\proba(y)=e^{-y}\theta(y)$; the points
$x_i(t+1)$ are then given (in an arbitrary order) by
\begin{equation}
x_i(t+1)=X_t+z+y_i,
\label{xit+1}
\end{equation}
and one gets from \eqref{defXt}
\begin{equation}
X_{t+1} = X_t + z + \ln\bigg[\sum_{i=1}^N e^{y_i}\bigg].
\label{Xt+1}
\end{equation}
We see that, with the definition (\ref{defXt}) of $X_t$, the differences
$X_{t+1}- X_t$ are independent variables. Therefore from
(\ref{LD})
\begin{equation}
e^{t G(\beta)} = \left\langle e^{-\beta (X_t-X_0)} \right\rangle =
\left\langle e^{-\beta (X_{t+1}- X_t)} \right\rangle^t ,
\end{equation}
and $G(\beta)$ can be computed by averaging over a single generation
\begin{equation}
\begin{aligned}
e^{G(\beta)}&=\langle e^{-\beta z}\rangle\bigg\langle \bigg[\sum_{i=1}^N
e^{y_i}\bigg]^{-\beta}\bigg\rangle\\
&= \frac{\Gamma(N+1+\beta)}{\Gamma(N+1)}
\int_0^\infty\diffd y_1 \cdots\int_0^\infty \diffd y_n\ 
e^{-y_1-\cdots-y_N} \bigg[\sum_{i=1}^N
e^{y_i}\bigg]^{-\beta}.
\end{aligned}
\end{equation}
Using the integral representation (valid for $\beta>0$)
\begin{equation}
A^{-\beta}
= \frac1{\Gamma(\beta)}\int_0^\infty\diffd\lambda\,
	\lambda^{\beta-1} e^{-\lambda A}
\qquad\text{(for $\beta>0$),}
\end{equation}
one obtains
\begin{equation}
e^{G(\beta)}=\frac{\Gamma(N+1+\beta)}{\Gamma(N+1)\Gamma(\beta)}
\int_0^\infty\diffd\lambda\,\lambda^{\beta-1} I_0(\lambda)^N,
\label{eGbeta}
\end{equation}
where the integral $I_0(\lambda)$ and more general integrals $I_p(\lambda)$ are defined as
\begin{equation}
I_p(\lambda)= \int_0^\infty \diffd y\, e^{(p-1)y-\lambda e^y}
=\lambda^{1-p}\int_\lambda^\infty \diffd u\, u^{p-2} e^{-u}.
\end{equation}
(These integrals are in fact, up to a simple change of variables,
incomplete gamma functions). For large $N$, the expression (\ref{eGbeta})
is dominated by small values of $\lambda$ where the integrals
$I_p(\lambda)$ for non-negative integers $p$ can be approximated by
\cite{BDMM.07}:
\begin{equation}
\begin{gathered}
I_0(\lambda)= 1+\lambda(\ln\lambda+\gamma_E-1)+{\cal O}(\lambda^2),
\qquad
I_1(\lambda)= -(\ln\lambda+\gamma_E)+{\cal O}(\lambda),
\\
I_{p\ge2}(\lambda)= {(p-2)!\over\lambda^{p-1}}+{\cal O}(\lambda^{2-p}),
\end{gathered}
\label{Ipsmall}
\end{equation}
where $\gamma_E=-\Gamma'(1) \simeq .577$ is Euler's constant.

Given the small $\lambda$ expansion of $I_0(\lambda)$, the integral in
\eqref{eGbeta} is dominated, for $N$ large and $\beta$ of order~1, by
$\lambda$ of order $1/(N\ln N)$. Making the change of variable $\lambda
= \mu/(N\ln N)$ one has
\begin{equation}
I_0(\lambda)^N = 
e^{-\mu}\left(1+\mu{\ln\mu-\ln\ln N+\gamma_E-1\over\ln N}+
{\cal O}\Big( \frac{\mu \ln \mu}{\ln N} \Big)^2
\right),
\label{I0^N}
\end{equation}
We can now evaluate
\eqref{eGbeta} for $N$ large and $\beta$ of order~1; using \eqref{I0^N} and
$\Gamma(N+1+\beta)/\Gamma(N+1)\simeq N^\beta$, one gets \cite{BDMM.07}
\begin{equation}
e^{G(\beta)}=\frac1{\ln^\beta N}
\left[1+\frac\beta{\ln N}\left({\Gamma'(\beta+1)\over\Gamma(\beta+1)}
-\ln\ln N+\gamma_E-1\right)+\cdots\right]
\end{equation}
or
\begin{equation}
G(\beta)=-\beta\ln\ln N + \frac\beta{\ln
N}\left({\Gamma'(\beta+1)\over\Gamma(\beta+1)}
-\ln\ln N+\gamma_E-1\right)+\cdots
\end{equation}
We see that, as $v=-G'(\beta)$, see (\ref{vbeta}), varying $\beta$ does not
change the leading $N$ dependence $v \simeq \ln \ln N$ of the velocity
but only shifts it by a small amount of order $\ln\ln N/\ln N$ which
vanishes in the $N\to\infty$ limit. We are now going to show that, on the
contrary, $\beta $ does change the statistical properties of the trees
even in the $N\to\infty$ limit.

\subsection{Trees}

We have already seen that all the offspring produced by generation $t$ are
distributed according to a Poisson point process of density
$\exp[{-(x-X_t)}]$. On the other hand, the offspring of
individual $x_i(t)$ are distributed as a Poisson point process of density
$\exp\big[{-[x-x_i(t)]}\big]$. This implies that, given that
there is an offspring in an interval $\diffd x$ around $x$, its probability
of being an offspring of $x_i(t)$ is 
\begin{equation}
W_i = e^{x_i(t)-X_t}=\frac{e^{x_i(t)}}{\sum_{j=1}^N e^{x_j(t)}}.
\end{equation}
This probability is independent of $x$. Therefore the probability $Q_p(t)$
that $p$ individuals at generation $t+1$ have the same ancestor at
generation $t$ is
\begin{equation}
Q_p(t)= \sum_{i=1}^N W_i^p
= \sum_{i=1}^N e^{p x_i(t)-p X_t}.
\end{equation}
If one weights these coalescence rates with the factor $e^{-\beta
X_{t}}$, then using \eqref{xit+1} and \eqref{Xt+1} with $t$ replaced by 
$t-1$ and using the fact that $X_{t-1}$, $z$ and the $y_i$ are independent,
one gets
\begin{equation}
\langle Q_p \rangle_\beta
 =N \frac{\big\langle e^{p x_1(t)-(\beta+p)X_t}\big\rangle}
 {\big\langle e^{-\beta X_t}\big\rangle}
 =N \frac{\Big\langle e^{p
y_1-(\beta+p)\ln\big[\sum_{i=1}^N e^{y_i}\big]}\Big\rangle}{\Big\langle
e^{-\beta\ln\big[\sum_{i=1}^N e^{y_i}\big]}\Big\rangle}
\end{equation}
with the $y_i$ independent exponential variables. The numerator and the
denominator can be computed in the same way as in \eqref{eGbeta} and one obtains
\begin{equation}
\langle Q_p \rangle_\beta
= N \frac{\Gamma(\beta)}{\Gamma(\beta+p)}\times\frac
{\int_0^\infty \diffd\lambda\,\lambda^{\beta+p-1} I_p(\lambda)
I_0(\lambda)^{N-1}}
{\int_0^\infty \diffd\lambda\,\lambda^{\beta-1} I_0(\lambda)^{N}}.
\end{equation}
We take $p\ge2$ and $\beta$ of order~1. The
integrals are dominated by $\lambda=\mu/(N\ln N)$ and $\mu$ of order~1. To
the leading order, $I_0(\lambda)^N\approx e^{-\mu}$, see \eqref{I0^N}, and
using \eqref{Ipsmall} for $I_p(\lambda)$ one easily gets, to leading order
\begin{equation}
\langle Q_p \rangle_\beta
\simeq\frac1{\ln
N}\,\frac{(p-2)!\,\Gamma(\beta+1)}{\Gamma(\beta+p)}
=\frac1{\ln N}\,\frac{(p-2)!}{(1+\beta)(2+\beta)\cdots(p-1+\beta)}.
\end{equation}
After rescaling time by a factor $\ln N$, one gets a coalescent with
transition rates $q_p=\langle Q_p\rangle_\beta \ln N$. One can check that
\begin{equation}
q_p= \frac{(p-2)!\,\Gamma(\beta+1)}{\Gamma(\beta+p)}=\int_0^1 
x^{p-2}\Lambda(\diffd x)\qquad\text{with }\Lambda(\diffd
x)=(1-x)^\beta\,\diffd x,
\label{qbeta}
\end{equation}
Using the expressions (\ref{T2T3T4}) of the appendix, where the ratios
${\langle T_3 \rangle / \langle T_2 \rangle}$ and ${\langle T_4 \rangle
/ \langle T_2 \rangle}$ have been obtained for a general coalescent,
one finally gets (\ref{beta}).

\section{The phenomenological theory}\label{sec:phenom}

In this section, we show that the phenomenological theory
developed in \cite{BDMM.06, BDMM.07} in the context of the noisy
Fisher-KPP equation predicts that \eqref{beta} remains valid for other
versions of the model described in the introduction. When the number of
offspring of each individual is bounded and when the shifts
$\epsilon_{i,j}$ are also bounded, one can describe the evolution of the
population by a noisy traveling wave equation of the Fisher-KPP type.
In \cite{BDMM.06, BDMM.07}, a phenomenological theory was proposed to
describe the large~$N$ behavior of these noisy equations. In the
$N\to\infty$ limit, the effect of noise vanishes and the traveling wave
has a finite velocity $v_\infty$ (in contrast to the exponential model
where the velocity diverges as $N\to\infty$). The
first correction when $N$ is large can be understood by
considering the cutoff introduced by the discrete number of particles: this
leads to $v_\text{cutoff}=v_\infty-A/\ln^2 N$. The next order correction
leads to a positive term of order $\ln\ln N/\ln^3 N$ which can be
understood, as well as the fluctuations of $X_t$,
by the following phenomenological theory: the front has, most of the time,
the shape and the velocity predicted by the cutoff theory. However,
every (typically) $\ln^3 N$ time steps, a rare event occurs where some
particles escape significantly ahead of the front. When this happens, the
shape of the front is at first deformed, but it relaxes to its cutoff shape
after $\sim\ln^2 N$ time steps. The end result is a finite increase of the
position of the front \cite{BDMM.07}. It has been shown that
this phenomenological theory predicts genealogies described by the
Bolthausen-Snitzmann coalescent. We are now going to show that when
we condition on the velocity by using the weight $\exp(-\beta X_t)$, this
leads to \eqref{beta}.

We consider a time interval $\Delta t$ which is large
compared to $\ln^2 N$ but small compared to $\ln^3 N$. During this
interval, there is a small probability $p(\delta)\,\diffd\delta\,\Delta t$
that an event of size $\delta$ occurs. When this happens, the
front position increases (after relaxation) by $R(\delta)$. The time
interval $\ln^2N\ll \Delta t\ll\ln^3N$ is such that
each event has the time to relax during $\Delta t$ and that 
the probability that two events
occur during the same time interval is negligible.

With these notations, the position $X_t$ of the front evolves according to
\begin{equation}
X_{t+\Delta t} - X_t = \begin{cases} v_\text{cutoff} \Delta t +R(\delta) &
\text{proba. } p(\delta)\,\diffd \delta\, \Delta t,\\
v_\text{cutoff} \Delta t +0 &\text{proba. } 1-\Delta t\int \diffd \delta\, p(\delta).
\end{cases}
\label{maineq}
\end{equation}
We argued in \cite{BDMM.07} that, for large $\delta$,
\begin{equation}
p(\delta)\approx C_1 e^{-\gamma \delta},\quad R(\delta)\approx
\frac1\gamma\ln\left(1+C_2 \frac{\gamma^3e^{\gamma\delta}}{\ln^3
N}\right),\quad C_1 C_2\approx\pi^2\gamma v''(\gamma).
\label{eqpheno}
\end{equation}
The number $\gamma$ and the function $v(r)$ depend on the details of the
model. (For $N\to\infty$, $v(r)$ gives the velocity of a front
starting with the initial condition $e^{-r x}$. For a step initial
condition, the system moves at the velocity $v(\gamma)$ where $\gamma$ is
the value at which $v(r)$ reaches its minimum.)

Using \eqref{maineq} to compute $G(\beta)$ given by \eqref{LD}, one gets
\begin{equation}
G(\beta)=-\beta v_{\text{cutoff}}+\int \diffd \delta\,p(\delta)\left[e^{-\beta
R(\delta)}-1\right].
\end{equation}

Let us now weight all the events by the factor $\exp(-\beta X_t)$.
\eqref{maineq} becomes
\begin{equation}
X_{t+\Delta t} - X_t = \begin{cases} v_\text{cutoff}
\Delta t +R(\delta) &
\text{proba. } \frac1 {Z(\beta)} e^{-\beta[v_\text{cutoff} \Delta t +R(\delta)]}
p(\delta)\,\diffd\delta\, \Delta t,\\
v_\text{cutoff} \Delta t +0 &\text{proba. } \frac1 {Z(\beta)} e^{-\beta
v_\text{cutoff} \Delta t}
\big[1-\Delta t\int \diffd\delta\, p(\delta)\big].
\end{cases}
\label{maineq*}
\end{equation}
$Z(\beta)$ is such that the probabilities are normalized; clearly
$Z(\beta)=e^{\Delta t G(\beta)}$.

Now, we can try to determine 
the probability $Q_p\,\Delta t$ that the $p$ particles coalesce
into one during the time interval $\Delta t$. 
We argued in \cite{BDMM.07} that when a rare event of
size $\delta$ occurs, a fraction $f=1-e^{-\gamma R(\delta)}$ of the population
is replaced by the offspring of the single particle that originated the
event ($\gamma$ is the model specific number appearing in \eqref{eqpheno}).
When this happens, there is a probability $f^p$ that the $p$ particles
coalesce during that interval of time $\Delta t$. This leads to
\begin{equation}
Q_p = \int \diffd\delta\, p(\delta)
e^{-\beta R(\delta)}
\Big[1-e^{-\gamma R(\delta)}\Big]^p
\label{Qppheno}
\end{equation}
We can now use \eqref{eqpheno} in \eqref{Qppheno}. Rewriting the integral in
term of the variable $f=1-e^{-\gamma R(\delta)}$ (the fraction of the
population replaced by the offspring of an individual), one gets
$p(\delta)\,\diffd\delta=C_1C_2\gamma^2/\ln^3 N\,\diffd f/f^2$, so that
\begin{equation}
Q_p=\frac{C_1C_2\gamma^2}{\ln^3N}\int_0^1
f^{p-2}(1-f)^{\beta/\gamma}\,\diffd f,
\end{equation}
which is the same as \eqref{qbeta} up to a prefactor (which only changes
the time scale) and the fact that $\beta$ is replace by $\beta/\gamma$.
Therefore the phenomenological theory leads to the same coalescent
as in the exponential model
but
with a different time scale of order $\ln^3N$ instead of $\ln N$.

\section{Comparison with mean-field spin glasses and the Poisson-Dirichlet
distribution}\label{sec:poissondirichlet}
\subsection{Various ways of characterizing random trees}

There are several ways of characterizing the statistical properties of the
trees generated by some given coalescence rates. (We only consider here
the cases where the coalescence rates do not vary in time, where the
particles play symmetrical roles and where at most one coalescence event can
occur during an interval of time~$\diffd t$.)
\begin{itemize}
\item
One can specify the coalescence rates $q_p$ (which take the values
(\ref{qbeta}) for the models of evolution with selection that we consider
in this paper). In terms of these coalescence rates, Kingman's coalescent
corresponds to
\begin{equation}
q_2 \neq 0,\qquad q_p=0 \quad\text{for} \ p \geq 3, 
\label{K2}
\end{equation}
while the Bolthausen-Sznitman coalescent corresponds to
\begin{equation}
q_p ={q_2 \over p-1}.
\label{BS2}
\end{equation}
It is easy to see that the rates (\ref{qbeta}) interpolate between
(\ref{K2}) for $\beta=\infty$ and (\ref{BS2}) for $\beta=0$.
\item
One can alternatively specify all the ratios ${\langle T_p \rangle
/ \langle T_2 \rangle}$ . It is clear (see (\ref{Tk},\ref{T2T3T4}) in the
appendix) that the knowledge of the $q_p$ determines all these time ratios
and conversely that the knowledge of the time ratios allows one to
calculate all the ratios $q_p/q_2$.
\item
One can also characterize the trees by the partition of the population they
induce at a given time in the past: the population at a generation $t$ can
be decomposed into several $\tau$-families where, by definition of these
families, two individuals $i$ and $j$ belong to the same $\tau$-family if
the age of their most recent common ancestor is less than $\tau$ (i.e.
$\tau_{i,j}(t) < \tau$). One can then associate to this $\tau$-partition of
the population at generation $t$ the following numbers
\begin{equation}
Y_k^{(\tau)}(t)= \big\langle \theta(\tau- \tau_{i_1,\ldots,i_k}) \big\rangle_t
\end{equation}
where $\langle\cdot \rangle_t$ means an average over all the possible
choices of the $k$ individuals $i_1,\ldots, i_k$ at generation $t$. 

One can interpret these $Y_k^{(\tau)}(t)$ as the probability that $k$
individuals chosen at random in the population at generation~$t$ belong
to the same $\tau$-family. These $Y_k^{(\tau)}(t)$ fluctuate from
generation to generation and the expressions $ \langle Y_k^{(\tau)}
\rangle$ of their averages over $t$ can be computed in terms of the
coalescence rates $q_p$. In the appendix, they are given for $k=2,3,4$ by
the quantities $Z_{k\to1}(\tau)\equiv \langle Y_k^{(\tau)} \rangle$.

Knowing all the $\langle Y_k^{(\tau)} \rangle$ (even for a single $\tau$)
determines also in principle all the coalescence rates and therefore all
the statistical properties of the trees.

\end{itemize}

\subsection{Comparison with mean-field spin glasses}

Very much like in the coalescence problems discussed above, where all the
individuals at a generation $t$ can be grouped into $\tau$-families, one
can group the spin configurations of a spin glass model according to their
distances $d$ (or to their overlap$=1-d$) in phase space. One can then
define \cite{MezardPSTV.84}, for a given sample, the probability $Y_k(d)$
that $k$ configurations, at thermal equilibrium, have all their $k(k-1)/2$
mutual distances in phase space less than $d$.

One of the predictions \cite{MezardPSTV.84,MezardParisiVirasoro.87,
DerridaFlyvberg.87} of the
Parisi solution \cite{Parisi.79,Parisi.80,Parisi.83} of the Sherrington
Kirkpatrick model
\cite{SherringtonKirkpatrick.75,KirkpatrickSherrington.78} is that this
$Y_k(d)$ fluctuates with the spin glass sample even when the system size
becomes large. The Parisi theory predicts also all the statistical
properties of these $Y_k(d)$. For example 
\begin{equation}
\langle Y_k(d) \rangle = \lim_{n \to 0} y_k(n,\mu) = {\Gamma(k-\mu ) \over \Gamma(k) \ \Gamma(1-\mu) } 
\label{spinglass}
\end{equation}
where
according to the broken replica symmetry 
\begin{equation}
y_k(n,\mu) = { \Gamma(1-n) \ \Gamma(k-\mu ) \over \Gamma(k-n) \ \Gamma(1-\mu)}.
\label{spinglass-bis}
\end{equation}
In (\ref{spinglass}) all the dependence on the distance $d$, on the details
of the model, and on the parameters such as the temperature or the magnetic
field is through the parameter $\mu$. Formula (\ref{spinglass-bis}) follows
from a very simple replica calculation: assume that one has $n$ replicas
grouped into ${n / \mu}$ families of $\mu$ replicas, $y_k(n,\mu)$ is
simply the probability that $k$ replicas chosen at random among the $n$
replicas belong to the same family.

For typical samples one has to take the $n \to 0$ limit as in
(\ref{spinglass}) and the statistics of the $Y_k$ coincide with those of
the Bolthausen-Sznitman coalescent: one can check that the $Z_{k \to
1}(\tau) \equiv \langle Y_k^{(\tau)} \rangle $ obtained in (\ref{Zkk})
coincides with (\ref{spinglass}) by choosing $\mu=e^{-q_2 \tau}$ and the
$q_p$ given by \eqref{BS2}.

In the spin glass case, one can also weight the samples according to their
free energy (by weighting them by a factor $Z^n$ where $Z$ is the partition
function \cite{Kondor.83,CoolenPenneySherrington.93}). One then expects from the
replica theory \cite{CoolenPenneySherrington.93} that the statistics of the
$Y_k(d)$ be modified and that
\begin{equation}
\frac{\langle Z^n \ Y_k(d) \rangle} {\langle Z^n \rangle }= y_k(n,\mu).
\end{equation}

One can check easily from \eqref{Zkk} with $q_p$ given by \eqref{qbeta}
that there exists no choice of $n$ and $\mu$ as functions of
$\beta$ and $\tau$ such that $y_k(n,\mu)=Z_{k\to1}(\tau)$. Therefore,
although the statistical properties of the
trees in the model of evolution with selection and in the spin glass
problem are the same for typical samples, they become different when
one introduces a bias (related to the free energy in the spin glass
problem and to the speed of adaptation in the models of evolution with
selection).

\subsection{The Poisson-Dirichlet distribution}

The Poisson-Dirichlet distribution \cite{PitmanYor.97, Berestycki.09} is
a probability distribution of the partitions of a unit interval into
infinitely many subintervals. It is parametrized by two parameters
$\alpha$ and $\theta$. One way of defining the Poisson-Dirichlet
distribution is to consider an infinite sequence $z_1$, $z_2$, \ldots,
$z_n$, \ldots of independent numbers, each $z_n$ being distributed
according to a distribution 
\begin{equation}
P_n(z_n)= \frac{\Gamma(1 + \theta+ n\alpha-\alpha) }
{ \Gamma(1 - \alpha)  \Gamma(\theta + n \alpha)}
z_n^{-\alpha} (1-z_n)^{\theta+ n \alpha -1} 
\end{equation}
(which is a $\beta$ distribution). Then one considers a partition of the
unit interval into subintervals of lengths $W_1$, $W_2$, \ldots $W_n$,
\ldots with
\begin{equation}
W_1+W_2+ \cdots+ W_n+\cdots =1,
\end{equation}
where the $W_i$ are given by
\begin{equation}\begin{aligned}
&W_1=z_1, \\
& W_2=(1-z_1) z_2, \\
& \ldots \\
& W_n=(1-z_1)(1-z_2) \cdots (1-z_{n-1}) z_n, \\
& \ldots 
\end{aligned}\end{equation}
For such a partition one can introduce the quantities 
\begin{equation}
Y_k = \sum_i W_i^k,
\end{equation}
which represents the probability that $k$ points chosen at random on the
unit interval fall in the same subinterval. It is easy to check that when
one averages over the $z_i$, one gets
\begin{equation}
\langle Y_k \rangle_{\alpha,\theta} = \big\langle z_1^k \big\rangle +
\big\langle
(1-z_1)^k \big\rangle \langle Y_k \rangle_{\alpha,\alpha+ \theta},
\end{equation}
The solution of this recursion is
\begin{equation}
\langle Y_k \rangle_{\alpha,\theta} = \frac{ \Gamma(1+ \theta) \Gamma(k-
\alpha) }{ \Gamma(k + \theta)  \Gamma(1-\alpha)}.
\label{P-D}
\end{equation}
One can notice \cite{Derrida.97} that these expressions are identical to
the replica expressions (\ref{spinglass-bis}) of $y_k(n,\mu)$ when one
chooses $\theta=-n$ and $\alpha= \mu$. Therefore as soon as one
introduces the bias $\beta \neq 0$ the statistical properties
$\tau$-families of our models of evolution with selection differ from
those of the Poisson-Dirichlet distribution.

\section{Conclusion}

In this paper we have seen that, for a family of simple models of
evolution under selection, the statistics of the genalogies are modified
when conditioning on the speed of evolution (\ref{beta}). For one
particular version of these models, the exponential model, the trees can
be generated by a coalescent with modified rates (\ref{qbeta}).
Numerical simulations (figures~\ref{fig:T3} and~\ref{fig:T4}) 
and a phenomenological theory
(section~\ref{sec:phenom})
indicate a similar behavior of more generic versions
of the model.

Despite their simplicity, there is not yet a full theoretical
understanding of the models of evolution with selection we
consider here. The introduction of the bias opens new questions which
would be interesting to consider. For exemple, what is the effect of the
bias on the steady state density profile of the population along the fitness
axis, or on the distances between the rightmost points in the population?
Numerically, the Monte-Carlo approach we developed here should give a
rather powerful tool to study these questions accurately and to test more
precisely the phenomenological theory developped in
\cite{BDMM.06,BDMM.07}.

It would also be interesting to study the genealogies of other models of
evolution with selection\cite{Kessler.97} to test the genericity of our results.

\bigskip

We are very happy to dedicate this work to David Sherrington, on the occasion of his $70^{\rm th}$ birthday.

\appendix

\section{Coalescence times and sizes of families in the $\Lambda$ coalescent}

In this appendix we calculate a few simple properties of a continuous time
coalescent defined as follows: one starts with $N$ points and during every
infinitesimal time interval $\diffd t \ll 1$, every subset of $k$ points has
a probability $q_k\, \diffd t$ of coalescing into one point. It is
assumed that there is at most one coalescence event during a time
interval $\diffd t$.

This model is called the $\Lambda$-coalescent
\cite{Pitman.99, MohleSagitov.01, Schweinsberg.00,
BerestyckiBerestyckiSchweinsberg.10, Berestycki.09}.
The coalescence rates can be written \cite{Pitman.99} in terms of a
positive measure $\Lambda$ on the interval $(0,1)$
\begin{equation}
q_k= \int_0^1 x^{k-2} \Lambda(\diffd x),
\end{equation}
and, more generally \cite{Pitman.99}, the rate $\lambda_{b,k}$ at
which the $k\ge2$ first
points out of $b$ coalesce into one point (while the other $b-k$ points
remain single) is given by
\begin{equation}
\lambda_{b,k}= \int_0^1 x^{k-2} (1-x)^{b-k}\, \Lambda(\diffd x)
=\sum_{n=0}^{b-k} \frac{(b-k)!}{n!(b-k-n)!}(-1)^n q_{n+k}
.
\label{lambdabk}
\end{equation}
It is more convenient in the following to use of the quantities
$r_b(b')$, defined as the rate at which a set of $b$ points coalesce into
a set of $b'<b$  points. Clearly
\begin{equation}
r_b(b') = \frac{b!}{(b'-1)!(b-b'+1)!} \lambda_{b,b-b'+1}
\label{rbb}
\end{equation}
(This simply means that the total number of distinct points jumps from
$b$ to $b'$ with probability $r_b (b') \diffd t$ during an infinitesimal
time interval $\diffd t$. The binomial factor in~\eqref{rbb} comes from the number of
ways of choosing the $b-b'+1$ points which coalesce.)

As already noticed, the rates (\ref{qbeta}) correspond to $
\Lambda(\diffd x)= (1-x)^\beta \diffd x$.

By analyzing what happens during a time interval $\diffd t$ one
can then see that the age $T_b$ of the most recent common ancestor of
$b$ individuals chosen  at generation $t+\diffd t$ satisfies
\begin{equation}
T_b(t+\diffd t) = \left\{\begin{array}{lll} 
 \diffd t & {\rm with \ probability} & r_b( 1)\, \diffd t, 
 \\ \diffd t+ T_2(t) & & r_b (2)\, \diffd t ,
 \\ \ldots & & 
 \\ \diffd t+ T_{b-1}(t) & & r_b (b-1)\, \diffd t ,
\\
 \diffd t+ T_b(t) & & 1 - \sum_{b'=1}^{b-1} r_b (b')\, \diffd t .
\end{array}
\right.
\label{Tk}
\end{equation}
Therefore one can determine recursively the average coalescence times 
by writing that $\langle T_b(t+\diffd t) \rangle=\langle T_b(t)
\rangle$, which leads to
\begin{equation}
 \left[\sum_{b'=1}^{b-1} r_b (b')\right]
\langle{T_b} \rangle= 1 + \sum_{b'=2}^{b-1} r_b (b') \langle T_{b'} \rangle 
\end{equation}
As (\ref{lambdabk}, \ref{rbb}) imply that
\begin{equation}
\begin{aligned}
r_2(1) &= q_2,  \\
r_3(1) &= q_3,\quad & r_3(2) &= 3(q_2-q_3), \\
r_4(1) &= q_4,\quad & r_4(2) &= 4(q_3-q_4),\quad &r_4(3) &= 6(q_2-2q_3+q_4),
\end{aligned}
\label{q1234}
\end{equation}
one gets 
\begin{equation}
\begin{aligned}
\langle T_2 \rangle &= \frac{1 }{ q_2},  \qquad\qquad
\frac{\langle T_3 \rangle }{ \langle T_2 \rangle} = \frac{4 q_2 - 3 q_3
}{ 3 q_2 - 2 q_3}, \\ 
\frac{\langle T_4 \rangle }{ \langle T_2 \rangle} &= \frac{27 q_2^2 - 56
q_2 q_3 + 28 q_3^2 + 12 q_2 q_4 - 10 q_3 q_4 }{ (3 q_2 - 2 q_3) ( 6 q_2 -
8 q_3 + 3 q_4)} .
\end{aligned}\label{T2T3T4}
\end{equation}

If one defines $Z_{b\to b'}(\tau)$ as the probability that $b$ points have
coalesced into $b'$ points during some time $\tau $, one can easily see
that it evolves according to
\begin{equation}
\frac {\diffd Z_{b\to b'} }{ \diffd \tau } 
= \sum_{b''> b'} r_{b''}( b') \ Z_{b\to b''} \ -\sum_{b''< b'} r_{b'}
(b'') \ Z_{b\to b'},
\end{equation}
with the initial condition that $Z_{b\to b'}(0)=\delta_{b,b'}$.
Then, using the expressions (\ref{q1234}) one gets 
\begin{eqnarray}
&& Z_{2\to2}= e^{- q_2 \tau },
\nonumber \\
&& Z_{2\to 1}=1-e^{- q_2 \tau },
\nonumber \\[2ex]
&& Z_{3\to 3}= e^{-(3 q_2-2 q_3) \tau },
\nonumber \\
&& Z_{3\to 2}= {3 \over 2} e^{-q_2 \tau } - {3 \over 2} e^{-(3 q_2-2 q_3)
\tau },
\nonumber \\
&& Z_{3\to 1}= 1- {3 \over 2} e^{-q_2 \tau } + {1 \over 2} e^{-(3 q_2-2
q_3) \tau },
\label{Zkk}\\[2ex]
&& Z_{4\to 4}= e^{ - (6q_2 - 8 q_3 + 3 q_4) \tau },
\nonumber \\
&& Z_{4\to 3}= 2 e^{ - (3q_2 - 2 q_3 ) \tau } -2 e^{ - (6q_2 - 8 q_3 + 3
q_4) \tau },
\nonumber \\
&& Z_{4\to 2}=
 {9q_2-14 q_3+ 5 q_4 \over 5 q_2 - 8 q_3+ 3 q_4} e^{-q_2 \tau } - 2 e^{ -
 (3q_2 - 2 q_3 ) \tau } +{6 q_2-10 q_3 + 4 q_4 \over 5 q_2 - 8 q_3+ 3
 q_4} e^{ - (6q_2 - 8 q_3 + 3 q_4) \tau },
\nonumber \\
&& Z_{4\to 1}=1 - {9q_2-14 q_3+ 5 q_4 \over 5 q_2 - 8 q_3+ 3 q_4} e^{-q_2
\tau } + e^{ - (3q_2 - 2 q_3 ) \tau } -{q_2-2 q_3 + q_4 \over 5 q_2 - 8
q_3+ 3 q_4} e^{ - (6q_2 - 8 q_3 + 3 q_4) \tau }.
\nonumber
\end{eqnarray}

\end{document}